\documentclass[conference]{IEEEtran}
\usepackage{cite}
\usepackage{booktabs}
\usepackage{array}
\usepackage{multirow}
\usepackage{pifont}
\usepackage{url}
\usepackage[hidelinks]{hyperref}
\usepackage{amsmath,amssymb}
\usepackage{graphicx}
\usepackage{xcolor}
\usepackage{tikz}
\usetikzlibrary{arrows.meta,positioning,fit,backgrounds}

\newcommand{\cmark}{\ding{51}}
\newcommand{\xmark}{\ding{55}}
\newcommand{\pmark}{$\sim$}
\newcommand{\ci}[2]{\,[#1, #2]}

\graphicspath{{figures/}{screens/}}

\begin{document}

\title{An Offline Smartphone Check for Drug Interactions and Past Reactions in Handwritten Prescriptions: A Technical Feasibility Study in Rural Bangladeshi Outpatient Care}


\author{\IEEEauthorblockN{Shahir Abdullah\textsuperscript{1}, Khondaker A. Mamun\textsuperscript{2,3}}
\IEEEauthorblockA{\textsuperscript{1}Department of Computer Science and Engineering\\
United International University, Dhaka, Bangladesh\\
\textsuperscript{2}CMED Health Limited (A Research Commercialization of AIMS Lab, UIU), Dhaka, Bangladesh\\
\textsuperscript{3}AIMS Lab, Department of CSE, United International University (UIU), Dhaka, Bangladesh\\
\textsuperscript{1}sshahnoor2620025@mscse.uiu.ac.bd, \textsuperscript{2,3}mamun@cse.uiu.ac.bd}}

\maketitle

\begin{abstract}
Rural Bangladeshi outpatient clinics keep no patient records, prescriptions are handwritten, and physicians have under a minute per patient, so they easily miss a known adverse drug reaction (ADR) or a documented drug--drug interaction (DDI). We built and evaluated an offline Android application that photographs a handwritten prescription, detects and reads each medicine line, and resolves brand names to active ingredients. The application \emph{retrieves}, and never predicts, documented interactions and the patient's recorded reactions. This quantitative technical-feasibility study, with a qualitative error analysis, involved no patients or clinicians. The reference data combine two Bangladeshi drug registries (27{,}880 brands) with 190{,}421 DDI pairs. A YOLOv8n line detector reached mAP@0.5 = 0.687 on a held-out prescription split but recalled only 60.6\% of lines at the deployed cutoff. On-device ML Kit text recognition read 11.8\% of 780 benchmark handwritten words exactly. We trained a CRNN-CTC reader on public handwriting data, decoded it against the brand lexicon, and then fine-tuned it on 660 author-labelled prescription crops. On benchmark words, the reader accepted the correct ingredient for 41.5\% and a wrong ingredient for 4.2\%; ML Kit at its preset cutoff accepted 21.0\% correctly and 5.3\% wrongly (McNemar $p<10^{-20}$). On scorable lines from held-out real prescriptions, the reader accepted the correct ingredient for 36.0\% [26.8, 46.3] and a wrong ingredient for 2.2\% [0.6, 7.8]. In 62 simulated runs on a mid-range phone, median time per patient was 26--31\,s and 59 of 62 runs finished within 60\,s, but 62\% of medicine lines still required a manual check. After the evaluation, we added a cross-visit medication history that checks a returning patient's new prescription against medicines resolved at their visits in the previous 30 days. It passed seven automated tests and one two-visit functional test but was not part of the evaluated runs. The pipeline is feasible within the clinic workflow; reading handwriting, not retrieval, is the bottleneck. Nothing here demonstrates a clinical benefit.
\end{abstract}


\section{Introduction}
Adverse drug reactions (ADRs) harm patients and burden health systems worldwide, and much of that harm is preventable~\cite{who2017}. Reviews estimate that ADRs cause roughly 5--6\% of hospital admissions in developed and developing countries alike, with most classified as preventable~\cite{angamo2016}. The harm extends to ambulatory care, where many adverse drug events are preventable or ameliorable~\cite{gurwitz2003}, and meta-analytic evidence suggests that about half of outpatient ADRs are avoidable~\cite{hakkarainen2012}. A recurrent, avoidable cause is giving a drug to a patient who has reacted to it before, or co-prescribing two drugs with a documented interaction.

In high-income systems this failure is mitigated by electronic health records (EHRs) that store a structured allergy history and raise an alert when a contraindicated drug is ordered. These systems suffer from \emph{alert fatigue}: clinicians override the large majority of drug-allergy alerts. At two academic hospitals, override rates exceeded 85\% over a decade, largely because the alerts were excessive and non-specific~\cite{topaz2016}. The consensus remedy is to make alerts more specific, so that warnings fire mainly for clinically significant events~\cite{poly2018}.

These safeguards are absent precisely where the harm is least tolerable. In Bangladesh, handwritten prescriptions dominate outpatient care, frequently contain omissions, and are often illegible. A survey of 900 prescriptions found illegible handwriting in nearly half~\cite{haque2016}, and prescribing errors remain common in rural district hospitals~\cite{sarkar2022}. Field observation at an upazila health complex in Meherpur indicated that the complex retains no structured record for non-admitted outpatients, that the registration identifier is a per-day serial that cannot link visits, and that a single duty physician may see about one patient per minute.

Digital clinical decision support (CDS) has proven feasible in comparable settings, typically as lightweight, purpose-built tools~\cite{muhindo2021,matthys2025,kallander2013}, but tools that match rural workflow poorly lose effectiveness~\cite{ciecierski2022}. Prior Bangladeshi studies have measured the prescription-error problem~\cite{haque2016,sarkar2022} but have not delivered a point-of-care medication-safety tool.

An earlier version of this work proposed such a tool and an evaluation plan. Here we report what we built and measured, including the dead ends. We make four contributions:
\begin{enumerate}
\item An \textbf{offline Android implementation} of the full pipeline. The app photographs a prescription, detects the medicine lines, asks the doctor to confirm them, reads each one, resolves brand names to active ingredients, and runs retrieval-only allergy and DDI checks. Every model runs on-device on a mid-range phone, and a 30-day \emph{cross-visit medication history} covers returning patients.
\item A \textbf{reference-data build} that merges two national registries and canonicalises ingredient names, so that the dictionary and interaction table share one vocabulary; this raises name coverage from 39.3\% to 60.6\%.
\item A \textbf{handwriting reader} for medicine names that limits its output to registered brands (lexicon-constrained CTC decoding), scores confidence at the ingredient level, and applies a three-tier accept / confirm / manual policy.
\item A \textbf{component-level and simulated-use evaluation} that keeps datasets strictly separate, reports confidence intervals and paired significance tests, and documents our protocol deviations.
\end{enumerate}
We claim no clinical-outcome effect: the study involved no patients, clinicians, or site records.
\section{Related Work}
\subsection{ADR burden, decision support and alert fatigue}
ADRs are frequent, costly and substantially preventable~\cite{who2017,angamo2016,gurwitz2003,hakkarainen2012}, with comparatively sparse evidence from low- and middle-income countries~\cite{angamo2016}. EHR-based allergy and interaction alerting is mature in high-income systems but undermined by very high override rates~\cite{topaz2016}; improving alert specificity is the key lever~\cite{poly2018}.

\subsection{CDS and mHealth in low-resource settings}
Lightweight phone- or tablet-based CDS has demonstrated feasibility and acceptability in rural LMIC facilities~\cite{muhindo2021,matthys2025,duffy2026guatemala}, and mobile phones are an accepted platform for frontline workers~\cite{kallander2013}. These studies target neonatal care, primary care or hypertension rather than medication-reaction safety, and workflow misfit is a known cause of failure~\cite{ciecierski2022}.

\subsection{Reading handwritten prescriptions}
Text-line recognition is commonly framed as sequence labelling with a convolutional--recurrent network trained with connectionist temporal classification (CTC)~\cite{shi2017crnn,graves2006ctc}; single-stage detectors such as YOLO~\cite{redmon2016yolo,jocher2023yolov8} localise text regions. General-purpose on-device recognisers (e.g.\ ML Kit~\cite{mlkit}) target printed text. Public Bangladeshi handwritten-prescription word datasets exist~\cite{ds_rxhandbd,ds_d4}, but we found no work that links reading to ingredient resolution and a safety check, or that reports what fraction of real prescription lines can be read \emph{safely}.

\subsection{Summary of the gap}
Table~\ref{tab:comparison} contrasts the main approaches against the requirements of the target setting.

\begin{table*}[t]
\caption{Comparison of existing approaches against the requirements of the target setting (rural Bangladeshi outpatient care).}
\label{tab:comparison}
\centering
\renewcommand{\arraystretch}{1.15}
\begin{tabular}{@{}p{3.6cm}ccccp{5.0cm}@{}}
\toprule
\textbf{Approach} & \textbf{ADR/DDI} & \textbf{LMIC fit} & \textbf{No-records infra.} & \textbf{Rural BD} & \textbf{Main limitation for our setting} \\
\midrule
EHR allergy/interaction alerts (high-income)~\cite{topaz2016,poly2018} & \cmark & \xmark & \xmark & \xmark & Requires full EHR/CPOE; severe alert fatigue \\
LMIC point-of-care CDS / mHealth~\cite{muhindo2021,matthys2025,kallander2013} & \xmark & \cmark & \cmark & \xmark & Targets other conditions; not medication safety \\
Bangladeshi prescription-error studies~\cite{haque2016,sarkar2022} & \pmark & \cmark & \pmark & \cmark & Measure the problem; no deployed system \\
Handwritten-prescription reading / OCR~\cite{ds_rxhandbd,ds_d4} & \xmark & \cmark & \pmark & \cmark & Reading only; no ingredient resolution, history or check \\
\textbf{This work} & \cmark & \cmark & \cmark & \cmark & Simulated use only; reading accuracy is the bottleneck \\
\bottomrule
\end{tabular}
\end{table*}

\section{Study Design}
\label{sec:design}
\subsection{Type of study}
This is a \textbf{quantitative technical-feasibility study with a qualitative error analysis}. The quantitative part measures each pipeline component on held-out public data and on author-labelled real prescriptions, and it measures end-to-end time per patient from the application's own logs. The qualitative part is a structured review of 60 screenshots and individual reading failures. The study involved \textbf{no human participants} other than the author, who acted as the simulated physician. It therefore yields no usability, acceptability, or interview data, and no clinical-outcome data.

\subsection{Research questions}
\begin{itemize}
\item \textbf{RQ1 (workflow):} Can the application process one patient in under a minute, offline, on a mid-range phone?
\item \textbf{RQ2 (components):} How accurately does the pipeline locate and read medicine lines, and how often does it accept a \emph{wrong} ingredient?
\item \textbf{RQ3 (retrieval):} Do the resolved ingredients reach the interaction knowledge base and the reaction history?
\end{itemize}

\subsection{Planned versus executed evaluation}
Table~\ref{tab:plan} states which parts of the evaluation plan in our earlier version were executed. We report this explicitly because the unexecuted parts are the ones that would establish clinical value.

\begin{table}[t]
\caption{Evaluation plan of the earlier version versus what was executed.}
\label{tab:plan}
\centering
\renewcommand{\arraystretch}{1.1}
\begin{tabular}{@{}p{2.6cm}p{5.1cm}@{}}
\toprule
\textbf{Planned axis} & \textbf{Status in this paper} \\
\midrule
Workflow fit (s/patient) & \textbf{Executed} as simulated use by the author: 62 logged runs, two sessions (Sec.~\ref{sec:res-time}). \\
Component accuracy & \textbf{Executed}: detection, reading, resolution, with CIs (Secs.~\ref{sec:res-det}--\ref{sec:res-rx3}). \\
Retrospective detection of known site cases & \textbf{Not executed}: no site records were used. \\
Identity re-linking via phone number & \textbf{Not executed}: dummy numbers only. \\
Usability / acceptability (e.g.\ SUS~\cite{brooke1996sus}, interviews) & \textbf{Not executed}: no clinicians took part. \\
Reaction-note reading (history capture) & \textbf{Implemented, not tested}: no reaction note was marked in any logged run. \\
Cross-visit medication history (30 days) & \textbf{Implemented after the evaluation; functionally tested only}: 7 automated tests and one two-visit run (Sec.~\ref{sec:crossvisit}); not part of the 62 evaluated runs. \\
\bottomrule
\end{tabular}
\end{table}

\section{Data}
\label{sec:data}
\subsection{Datasets and their roles}
We use five data sources, (Table~\ref{tab:data}). We take box-location metrics only from Dataset~3 and word-reading benchmarks only from Dataset~4, and we never blend the two.

\begin{table*}[t]
\caption{Data sources and their roles. Sources marked ``Never trained on'' are lookup tables. We never report results on a split used for model selection.}
\label{tab:data}
\centering
\renewcommand{\arraystretch}{1.12}
\begin{tabular}{@{}p{0.4cm}p{4.1cm}p{4.5cm}p{7.2cm}@{}}
\toprule
\textbf{\#} & \textbf{Source} & \textbf{Content} & \textbf{Role} \\
\midrule
1 & BdMedex~\cite{medex} + DGDA registered products~\cite{dgda} & 21{,}714 + 41{,}201 brand rows & Reference: brand-to-ingredient lookup. Never trained on. \\
2 & DrugBank-derived DDI table~\cite{ds_ddi,huang2021tdc, wishart2018drugbank} & 191{,}541 drug pairs with text & Reference: interaction knowledge base. Never trained on. \\
3 & Handwritten prescriptions with medicine-line boxes~\cite{ds_rxboxes} & 200 photos, 1{,}086 boxes (YOLO format) & Detector: train 140 / val 30 / test 30 photos (seed 42). The author also added text labels to the boxes (Sec.~\ref{sec:labels}). \\
4 & Doctor's Handwritten Prescription BD~\cite{ds_d4} & 4{,}680 word images, 78 names (+generic) & \textbf{OCR benchmark only}: test split, 780 images. Never trained or tuned on. \\
5 & RxHandBD~\cite{ds_rxhandbd} & 5{,}578 word images, 1{,}509 distinct texts & We train the handwriting reader on 4{,}463 images and use the test split (1{,}115) to select checkpoints and the first cutoff. No image appears in both this dataset and Dataset~4, but 33 of Dataset~4's 78 names appear in Dataset~5's training labels. \\
\bottomrule
\end{tabular}
\end{table*}

Two further sources were inspected and \emph{not} used: a byte-identical copy of Dataset~5 (``RxHand Original''), and four non-allopathic product lists (ayurvedic, herbal, homeopathic, unani). A third DGDA file listing 1{,}499 generics without brands was not usable for brand mapping. A registry list with a ``use for'' field (\texttt{allopathic.json}, 35{,}015 rows, 1{,}888 veterinary) was used only to exclude veterinary products.

\subsection{Text labels for real prescription crops}
\label{sec:labels}
Dataset~3 has boxes but no text. To measure reading on \emph{real} prescription crops, every labelled box was cropped exactly as the application crops it (8\% horizontal and 16\% vertical padding) and the author typed its content in an offline labelling page. Of 1{,}086 boxes, 124 were marked unreadable (``?'') and 90 as not a medicine (``\#'', e.g.\ ``Endoscopy'', advice, Bangla text). The crops inherit the detector split: \textbf{train} crops fine-tune the reader, \textbf{val} crops (174) select its checkpoint and cutoff, and \textbf{test} crops (165, from the 30 test photos) are used only for reporting and were never uploaded to the training environment. Labels were produced by a single annotator (the author) without double coding; this is a limitation (Sec.~\ref{sec:threats}).

\subsection{Reference-data construction}
\label{sec:refdata}
\textbf{Dictionary.} From BdMedex, 351 herbal rows were dropped, leaving 21{,}363 allopathic rows and 13{,}611 distinct brand names; 93 brand names carried different generics across dosage forms, for which we keep the \emph{union} of ingredients (a missed ingredient is a missed flag; an extra one is at worst an over-flag). The DGDA export added 14{,}269 brands after removing 1{,}066 veterinary-only rows, 818 bulk/raw-material/bolus rows and 31 rows without a generic; 10{,}189 brands appear in both sources. The final dictionary has \textbf{27{,}880 brands} over 1{,}682 ingredients, of which 7{,}146 are fixed-dose combinations.

\begin{figure}[t]
\centering
\includegraphics[width=\columnwidth]{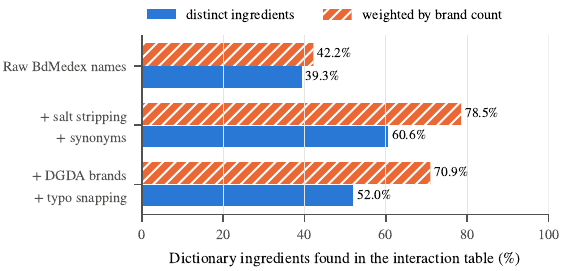}
\caption{Share of dictionary ingredients present in the DDI table at each construction stage. Adding the DGDA registry lowered the \emph{share} (it adds vitamin, mineral and vaccine combinations absent from DrugBank) while raising the absolute number of covered ingredients from 762 to 875.}
\label{fig:coverage}
\end{figure}

\textbf{Canonical ingredient names.} BdMedex names carry salt forms (``ceftriaxone sodium'') while the DDI table uses the active moiety (``ceftriaxone''). One shared function strips salt and hydrate words (except when the first word is itself an ion, e.g.\ ``calcium carbonate''), applies 27 reviewed synonyms mapping DDI names to local names (e.g.\ acetaminophen $\rightarrow$ paracetamol, cefradine $\rightarrow$ cephradine), and, for DGDA names absent from the known vocabulary, snaps typographical variants to the nearest known name at normalised Levenshtein similarity $\geq 0.88$~\cite{levenshtein1966}. All 48 proposed snaps were reviewed; three false ones (e.g.\ sodium carbonate $\rightarrow$ sodium bicarbonate; 5-aminosalicylic $\rightarrow$ 4-aminosalicylic acid) were blocked, leaving 45. Fig.~\ref{fig:coverage} shows the effect on the share of dictionary ingredients that exist in the DDI table.

\textbf{Interaction table.} The 191{,}541 DDI rows reduce to \textbf{190{,}421} unordered pairs over 1{,}697 canonical ingredients. The table has no severity field; consequently the system cannot rank interactions (Sec.~\ref{sec:qual}). The ``severe-reaction'' reference list of the earlier design contains only two placeholder entries; in practice \emph{any} reaction recorded for a patient is treated as flag-worthy.

\section{System}
\label{sec:system}
\subsection{Architecture and workflow}
Fig.~\ref{fig:arch} shows the implemented pipeline. Reception records a phone number (the only identifier); the physician writes as usual and photographs the prescription (or picks a scan). All computation runs on the phone; there is no server.

\begin{figure}[t]
\centering
\begin{tikzpicture}[font=\scriptsize, node distance=2.6mm,
  box/.style={draw, rounded corners=2pt, align=center, inner sep=2.5pt, text width=6.6cm, minimum height=5mm},
  hum/.style={box, fill=black!4}, ml/.style={box, fill=black!12}, ret/.style={box, fill=black!22},
  ar/.style={-{Stealth[length=1.6mm]}, semithick}]
\node[hum] (a) {Reception: patient phone number (soft identifier)};
\node[hum, below=of a] (b) {Physician writes; photographs prescription (camera or gallery)};
\node[ml, below=of b] (c) {Line detector (YOLOv8n, TFLite, $1024^2$ input, conf $\geq 0.15$)};
\node[hum, below=of c] (d) {Doctor confirms: tap to drop a box, drag to add a missed line or a reaction note};
\node[ml, below=of d] (e) {Reader per box: CRNN-CTC $\rightarrow$ lexicon CTC over 27{,}872 brands $\rightarrow$ ingredient-set posterior $P$};
\node[box, below=of e, fill=white] (f) {$P\geq0.995$: accept \quad $0.9\leq P<0.995$: ``Is this X?'' (Yes/No) \quad else: manual check};
\node[ret, below=of f] (g) {Retrieval: (A) ingredients $\cap$ patient's recorded reactions; (B) every ingredient pair $\in$ DDI table; (C) today's ingredients $\times$ medicines from this patient's visits in the last 30 days};
\node[hum, below=of g, xshift=0mm] (s) {Store this visit's \emph{resolved} medicines (auto-accepted or confirmed) under the phone number};
\node[hum, below=of s] (h) {Result: flags, or ``Good to go'' only if every line is resolved and nothing matched};
\foreach \u/\v in {a/b,b/c,c/d,d/e,e/f,f/g,g/s,s/h} \draw[ar] (\u) -- (\v);
\end{tikzpicture}
\caption{Implemented pipeline. Light boxes: human steps; mid-grey: on-device learned components; dark: retrieval. Reaction notes are read with ML Kit and stored at ingredient level against the phone number. Check (C) and the visit store were added after the evaluated sessions (Sec.~\ref{sec:crossvisit}).}
\label{fig:arch}
\end{figure}

\subsection{Implementation}
\begin{figure*}[t]
\centering
\includegraphics[width=0.235\textwidth]{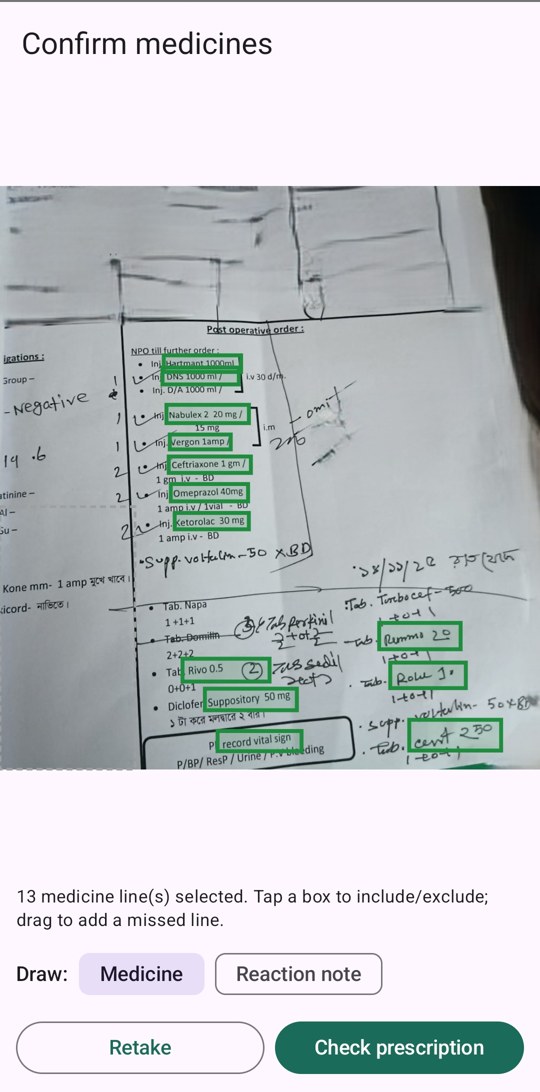}\hfill
\includegraphics[width=0.235\textwidth]{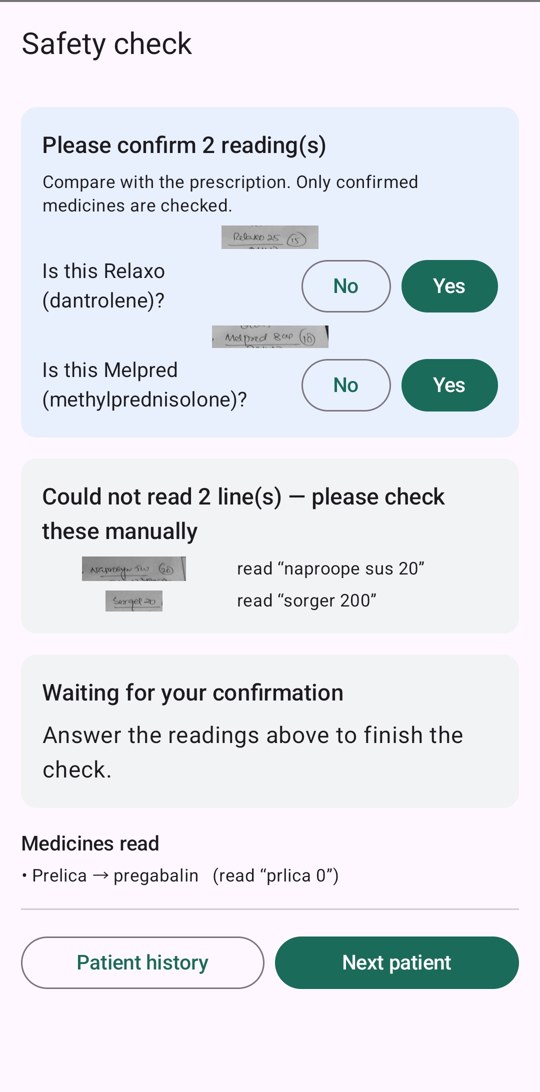}\hfill
\includegraphics[width=0.235\textwidth]{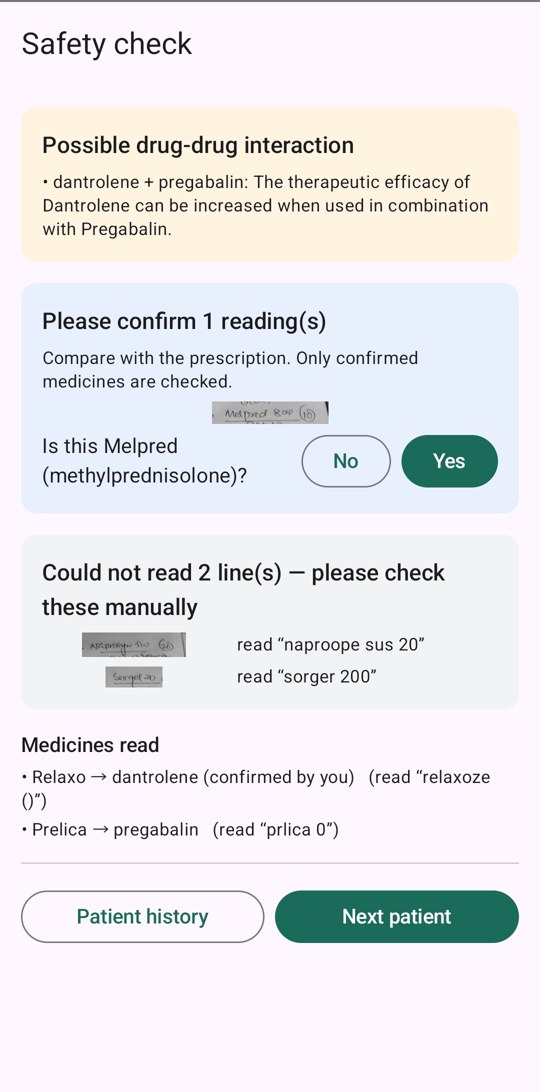}\hfill
\includegraphics[width=0.235\textwidth]{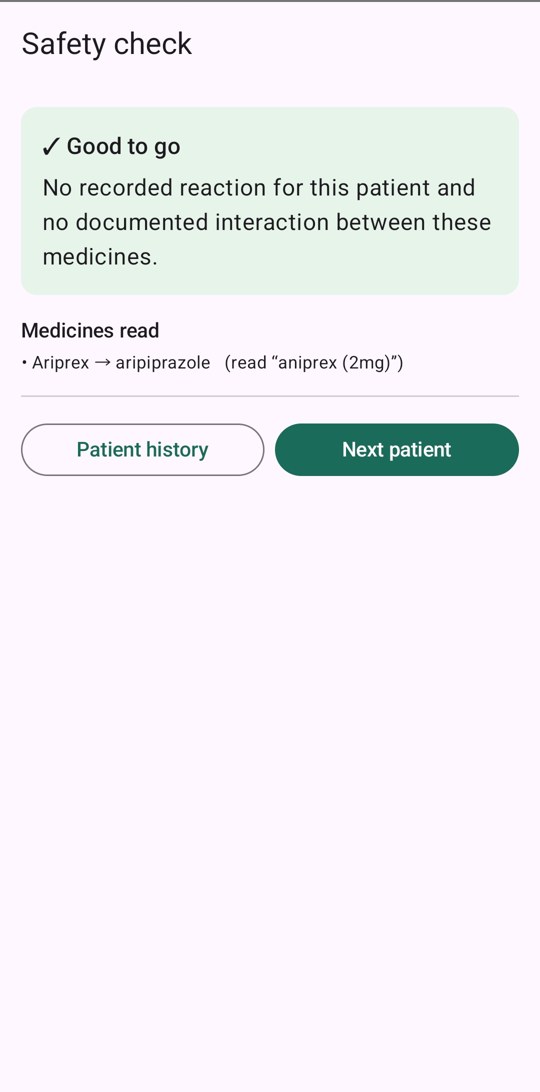}
\caption{Application screens from the second simulated-use session (test-split images of Dataset~3). (a) Detector suggestions for confirmation; (b) ``Is this X?'' confirmations and unreadable lines with their crops; (c) a DDI warning (dantrolene + pregabalin) that arose because the author confirmed ``Relaxo'' for a line actually labelled ``Rolac 25''---a wrong human confirmation (Sec.~\ref{sec:qual}); (d) ``Good to go''.}
\label{fig:screens}
\end{figure*}

The application is written in Kotlin with Jetpack Compose and Room (SQLite). Reference CSVs are bundled as assets (dictionary 1.5\,MB, DDI 27\,MB) and streamed into the database on first launch inside one transaction per table (22\,s on the test phone); a content fingerprint triggers a reload of reference tables after a data update while preserving patient history. The detector (12.4\,MB float32) and reader (7.4\,MB) run with LiteRT/TensorFlow Lite~\cite{litert}; reaction notes use ML Kit Latin text recognition~\cite{mlkit}. The debug APK is 99.8\,MB. The phone number can be typed or, holding a button, spoken in English; Android's on-device speech recogniser fills the field and the physician corrects it if needed. This option was added after the evaluated sessions and was not timed or evaluated. The test device was a Samsung Galaxy A52 (SM-A525F, Android~14, 256\,MB app heap). Every run writes a timing record (no patient data; random run identifier), exported as CSV for the analysis in Sec.~\ref{sec:res-time}. Fig.~\ref{fig:screens} shows the main screens.

\subsection{Line detector}
\begin{figure}[t]
\centering
\includegraphics[width=\columnwidth]{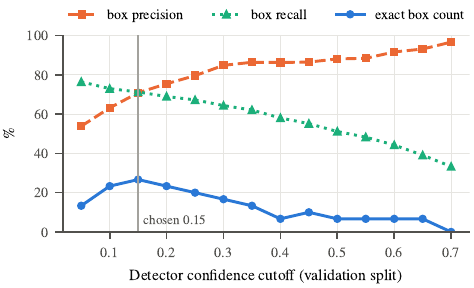}
\caption{Effect of the detector's confidence cutoff on the validation split. The deployed 0.15 trades precision for recall because a missed line is a missed drug.}
\label{fig:detconf}
\end{figure}

YOLOv8n~\cite{jocher2023yolov8} (3.0\,M parameters, 8.1\,GFLOPs) was fine-tuned from COCO weights on Dataset~3 for 100 epochs at $1024^2$ on a Colab T4 GPU (0.14\,h) with document-safe augmentation: no flips, rotation $\leq 5^\circ$, translation 0.05, scale 0.2, saturation/value jitter 0.3, mosaic 0.5. The raw labels used class id~1 with a single declared class; they were remapped to~0 (otherwise every box is discarded), and \texttt{.jpg}/\texttt{.png} images (12 of 200) that the scaffold skipped were included. The TFLite export was checked against the PyTorch model to $\pm1$\,px on a test image. The confidence cutoff (0.15) was chosen on the val split as the value maximising exact box count (Fig.~\ref{fig:detconf}).

\subsection{Handwriting reader and lexicon decoding}
\label{sec:reader}
The reader is a CRNN~\cite{shi2017crnn}: four conv--BN--ReLU blocks reduce a $32{\times}128$ greyscale strip to 64 time steps, followed by two bidirectional LSTMs (128 units) and a softmax over 45 characters plus the CTC blank~\cite{graves2006ctc} (1.65\,M parameters). Inputs are contrast-stretched (2nd--98th percentile), inverted and stretch-resized; the identical preprocessing is reimplemented in Kotlin. Training used Adam with cosine decay (lr $10^{-3}$, 120 epochs, batch 64) and on-GPU augmentation (rotation, zoom, shift, stroke dilation/erosion, contrast, random binarisation, noise). For on-device export the LSTMs are unrolled over the fixed 64 steps so that the model converts to built-in TFLite operations only.

Free-text output is not trusted. Instead, for every dictionary brand $b$ the CTC log-likelihood $\ell_b=\log p(b\mid x)$ of its name is computed (with an optional strength suffix taken from the greedy read). To make this fast on a phone, an order-free pre-score $\sum_{c\in b}\max_t \log p_t(c)$ shortlists 2{,}000 brands before exact scoring; on 195 benchmark images this shortlist always contained the full-lexicon winner, and latency fell from 943 to 160\,ms per line. The confidence used for acceptance is the \textbf{ingredient-set posterior}
\begin{equation}
P(I^\ast\mid x)=\frac{\sum_{b:\,\mathrm{ing}(b)=I^\ast} e^{\ell_b}}{\sum_{b} e^{\ell_b}},\qquad I^\ast=\mathrm{ing}\big(\arg\max_b \ell_b\big),
\end{equation}
which does not split probability mass between look-alike brands with the same ingredients (e.g.\ Rosumax/Rosuva). Readings with $P\geq\tau_a$ are accepted, those with $\tau_s\leq P<\tau_a$ are shown as ``Is this X?'' (never used unconfirmed), and the rest go to manual check, with $\tau_s=0.9$ and $\tau_a$ chosen as in Sec.~\ref{sec:protocol}.

\subsection{Safety logic}
All three checks are lookups. (A) The application intersects the patient's recorded reaction ingredients with the prescription's ingredients. (B) It looks up every unordered ingredient pair in the DDI table by a canonical key. (C) \emph{Cross-visit:} it pairs each ingredient prescribed today with every ingredient resolved at the patient's earlier visits in the last 30 days (using each ingredient's most recent occurrence) and looks up these pairs in the same table. Check C skips past ingredients that also appear on today's prescription, because check B already covers them.

After running the checks, the application stores the visit's \emph{resolved} medicines (auto-accepted lines and ``Is this X?'' lines confirmed with Yes) with the date under the patient's phone number. It never stores unreadable or rejected lines, so nothing guessed enters the history. The screen shows ``Good to go'' only when all three conditions hold: no check flagged anything, every medicine line was resolved, and no confirmation is pending. Otherwise, it names the lines that need a manual check.

\section{Development History}
\label{sec:history}
The system was not designed in one pass; Table~\ref{tab:iter} lists the iterations, the evidence that triggered each, and its effect, including mistakes that changed the results.

\begin{table*}[t]
\caption{Incremental development. ``D4'' = Dataset~4 benchmark (780 words, on-device); IC / WA = ingredient-correct / wrong-accepted (\% of all words) at the stated cutoff.}
\label{tab:iter}
\centering
\renewcommand{\arraystretch}{1.12}
\begin{tabular}{@{}p{0.3cm}p{3.6cm}p{6.1cm}p{6.6cm}@{}}
\toprule
\textbf{\#} & \textbf{Change} & \textbf{Trigger} & \textbf{Effect} \\
\midrule
1 & Ingredient canonicalisation (salt, synonyms) & Only 39.3\% of dictionary ingredients existed in the DDI table; same drug stored under two spellings & Coverage 60.6\%; one spelling per ingredient, so reactions match across brands \\
2 & Full-dictionary fuzzy match & Scaffold matched only the first 50 brands sharing a 3-letter prefix; OCR gets the first 3 letters right 23\% of the time & +1.7 pp ingredient-correct (EasyOCR proxy) \\
3 & ML Kit $\rightarrow$ first real test & Author's first photographs: ``Napa Extend'' read as ``e.\ Nopa Euknd'' & 3/3 lines unreadable; correct brand ranked 19th and 39th by fuzzy match \\
4 & DGDA merge, vet filter, typo snapping & Brands on real prescriptions missing from BdMedex & 13{,}611 $\rightarrow$ 27{,}880 brands \\
5 & CRNN-conv reader + lexicon CTC & Need a handwriting model; free text unrecoverable & D4: IC 30.0\% / WA 4.6\% @0.99 (cutoff read off D4, see \#8) \\
6 & CRNN-LSTM reader & LSTM export to TFLite initially failed; conv fallback overwrote one LSTM run & D4 greedy exact 15.1\% $\rightarrow$ 21.3\% \\
7 & 2{,}000-brand shortlist & 943\,ms per line on the phone & 160\,ms; accuracy unchanged (30.0\% vs 30.1\%) \\
8 & Cutoff chosen on RxHand test, not D4 & Earlier cutoffs had been read off the benchmark itself & LSTM: 0.9999 $\rightarrow$ IC 22.9\% / WA 2.1\% (honest) instead of 38.0\% / 4.5\% (optimistic) \\
9 & Ingredient-set posterior & Look-alike brands split confidence & IC 22.9\% $\rightarrow$ 26.8\% at the same WA \\
10 & ``Is this X?'' confirmation tier & Only 6\% of real crops reached $P\geq0.9999$ & 0.9--$\tau_a$ readings become one-tap decisions \\
11 & Fine-tuning on 660 real crops & Real-crop CER 0.63 vs 0.26 on RxHand: domain gap & Real-crop CER 0.40; D4 IC 41.5\% / WA 4.2\% @0.995 \\
12 & Cutoff cap 5\% $\rightarrow$ 2\% & 5\% cap on val gave 0.98, which accepted 7.9\% wrong on \emph{test} & $\tau_a=0.995$ (val WA 1.8\%). Cap changed after seeing test: disclosed, both points reported \\
13 & Cross-visit medication history (30 days) & Screenshot review: returning patients' earlier prescriptions were invisible to the check & Check (C) + visit store; 7/7 automated tests pass; functional demo flagged diclofenac (today) + pantoprazole (earlier visit) \\
\bottomrule
\end{tabular}
\end{table*}

\begin{figure}[t]
\centering
\includegraphics[width=\columnwidth]{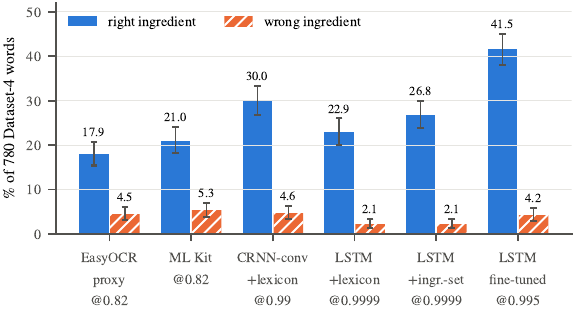}
\caption{Reading iterations on Dataset~4 (on-device; EasyOCR is a desktop proxy). Bars: ingredient-correct and wrong-accepted at the stated cutoff; whiskers: 95\% Wilson CIs. The conv model's cutoff (0.99) was read off this benchmark and is therefore optimistic; later cutoffs were chosen on other data.}
\label{fig:prog}
\end{figure}

Fig.~\ref{fig:prog} summarises the reading iterations on the Dataset~4 benchmark at each reader's operating point.

\section{Evaluation Protocol}
\label{sec:protocol}
\subsection{Metrics}
For a set of $N$ items (words or medicine lines), each item $i$ has a gold ingredient set $G_i$, a prediction $\hat I_i$, and a confidence $P_i$. At cutoff $\tau$, we compute two rates:
\begin{align}
\mathrm{IC}(\tau)&=\tfrac{1}{N}\textstyle\sum_i \mathbb{1}[P_i\geq\tau \wedge \hat I_i=G_i],\\
\mathrm{WA}(\tau)&=\tfrac{1}{N}\textstyle\sum_i \mathbb{1}[P_i\geq\tau \wedge \hat I_i\neq G_i].
\end{align}
Items below the cutoff go to manual review. WA is the safety-relevant error: a silently wrong ingredient can hide an interaction or invent one. Varying $\tau$ traces a risk--coverage curve~\cite{geifman2017selective}. We judge correctness at the ingredient level, so a different brand with the same ingredients counts as correct. For real crops, we obtain the gold ingredient set by stripping the strength from the typed label and looking the name up in the dictionary. Lines whose written name is not a dictionary brand are unscorable, and we report them separately. We evaluate detection with mAP and box precision and recall at IoU $\geq 0.5$, plus exact box count per prescription, and we score raw reading by exact match and character error rate (CER).

\subsection{Cutoff selection}
We set the accept cutoff $\tau_a$ to the lowest grid value whose WA on a \emph{selection} split is at most a cap, then apply it unchanged to reporting splits. For the RxHand-only reader, we used RxHandBD's test split, restricted to the 822 words that are dictionary brands, as the selection split; a 5\% cap gives $\tau_a=0.9999$. For the fine-tuned reader, we used the real-crop validation split (112 scorable lines); a 5\% cap gives 0.98, and a 2\% cap gives 0.995. We first used a 5\% cap, saw 7.9\% WA on the test split, and then adopted a 2\% cap. Because we changed the cap \emph{after} seeing test results, we report both operating points.

\subsection{Statistics}
We report 95\% Wilson score intervals for proportions~\cite{wilson1927}. For paired comparisons on the same items, we use the exact McNemar test~\cite{mcnemar1947} on the binary outcome ``accepted and correct''. We report 95\% bootstrap intervals for timing medians (10{,}000 resamples), compare the two sessions with the Mann--Whitney U test, and compare manual-check shares with Fisher's exact test. A script recomputes all numbers in this paper from saved artefacts; we transcribe none.

\section{Results}
\subsection{Line detection (Dataset 3, held-out 30 photos)}
\label{sec:res-det}
\begin{figure}[t]
\centering
\includegraphics[width=\columnwidth]{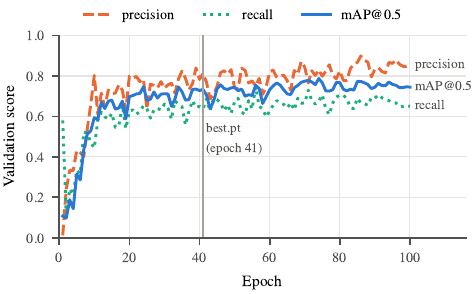}
\caption{Detector validation curves (30 photos). The checkpoint at epoch 41 maximised Ultralytics' fitness ($0.1\,\mathrm{mAP@0.5}+0.9\,\mathrm{mAP@0.5{:}0.95}$).}
\label{fig:dettrain}
\end{figure}

Table~\ref{tab:det} reports detection on the test split; Fig.~\ref{fig:dettrain} shows training. Precision is reasonable but recall is low: at the deployed cutoff 39\% of medicine lines are missed, and the box count is exactly right on only 5 of 30 prescriptions (16 under-counted, 9 over-counted). Many misses are short abbreviated names (e.g.\ ``AZ 500''). Part of the apparent error is label noise: some boxes in Dataset~3 mark tests (``Endoscopy'', ``Colonoscopy''), which the detector correctly ignores.

\begin{table}[t]
\caption{Detection on the Dataset~3 test split (30 photos, 165 boxes).}
\label{tab:det}
\centering
\begin{tabular}{@{}lr@{}}
\toprule
mAP@0.5 / mAP@0.5:0.95 & 0.687 / 0.314 \\
Precision / recall (at max-F1 confidence) & 0.845 / 0.564 \\
Box precision / recall at cutoff 0.15 (IoU $\geq$ 0.5) & 71.4\% / 60.6\% \\
Exact box count per photo & 16.7\%\ci{7.3}{33.6} \\
Mean absolute count error & 1.70 boxes \\
\bottomrule
\end{tabular}
\end{table}

\subsection{Reading on the Dataset 4 benchmark (on-device)}
\label{sec:res-d4}
\begin{figure}[t]
\centering
\includegraphics[width=\columnwidth]{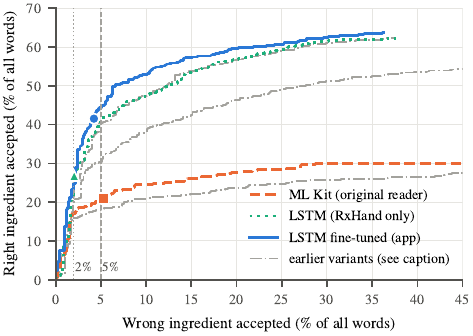}
\caption{Risk--coverage on Dataset~4 (on-device): each curve sweeps the acceptance cutoff; markers are the operating points of Table~\ref{tab:d4}. Grey dash-dot curves: EasyOCR proxy (lowest), CRNN-conv (middle) and the LSTM with brand posterior (nearly coincident with the ingredient-set LSTM). Dotted/dashed verticals: 2\% and 5\% wrong-accepted.}
\label{fig:d4}
\end{figure}

Table~\ref{tab:d4} and Fig.~\ref{fig:d4} compare readers on the 780 Dataset~4 test words, all measured on the phone except the EasyOCR proxy~\cite{easyocr}. ML Kit returned no text for 38.3\%\ci{35.0}{41.8} of words. The Kotlin reader produced the same greedy text as the Python reference on 778/780 images. At its operating point, the fine-tuned reader accepted the correct ingredient for 41.5\% of words versus 21.0\% for ML Kit; on the same images, 218 words were correct only with the fine-tuned reader and 58 only with ML Kit (exact McNemar $p=5.6\times10^{-23}$). Its WA (4.2\%) is lower than ML Kit's (5.3\%) but within sampling error.

\begin{table*}[t]
\caption{Dataset~4 test words ($N=780$), on-device. Exact = greedy text equals the name. IC/WA at the operating cutoff with 95\% Wilson CIs. ``Top-1'' = always accept the best lexicon/fuzzy match (no cutoff). $^\dagger$Cutoff read off this benchmark (optimistic). ms/line: median on the Galaxy A52 (ML Kit: recognition only; CRNN: including lexicon decoding).}
\label{tab:d4}
\centering
\renewcommand{\arraystretch}{1.12}
\begin{tabular}{@{}lcccccr@{}}
\toprule
\textbf{Reader} & \textbf{Exact} & \textbf{Cutoff} & \textbf{IC (\%)} & \textbf{WA (\%)} & \textbf{Top-1 IC (\%)} & \textbf{ms/line} \\
\midrule
EasyOCR + fuzzy (desktop proxy) & --- & 0.82 & 17.9\ci{15.4}{20.8} & 4.5\ci{3.2}{6.2} & 28.1 & --- \\
ML Kit + fuzzy (original) & 11.8\% & 0.82 & 21.0\ci{18.3}{24.0} & 5.3\ci{3.9}{7.1} & 30.0 & 128 \\
CRNN-conv + lexicon & 15.1\% & 0.99$^\dagger$ & 30.0\ci{26.9}{33.3} & 4.6\ci{3.4}{6.3} & 54.4 & 160 \\
CRNN-LSTM + lexicon (brand posterior) & 21.3\% & 0.9999 & 22.9\ci{20.1}{26.0} & 2.1\ci{1.3}{3.3} & 62.3 & 207 \\
CRNN-LSTM + lexicon (ingredient-set) & 21.3\% & 0.9999 & 26.8\ci{23.8}{30.0} & 2.1\ci{1.3}{3.3} & 62.3 & 198 \\
\textbf{Fine-tuned LSTM + lexicon (ingredient-set), deployed} & \textbf{24.5\%} & \textbf{0.995} & \textbf{41.5}\ci{38.1}{45.0} & \textbf{4.2}\ci{3.0}{5.9} & \textbf{63.7} & \textbf{200} \\
\bottomrule
\end{tabular}
\end{table*}

\textbf{Familiar names are read more confidently wrong.} Splitting Dataset~4 by whether the name occurs in RxHandBD's training labels, the deployed reader accepted the correct ingredient for 52.1\%\ci{46.7}{57.5} of the 330 words with \emph{seen} names but with 7.6\%\ci{5.2}{10.9} wrong, against 33.8\%\ci{29.6}{38.3} correct and 1.8\%\ci{0.9}{3.5} wrong for the 450 words with \emph{unseen} names. The reader tends to see a familiar training brand in similar handwriting; the overall WA on any benchmark therefore depends on its overlap with the training vocabulary.

\subsection{Reading on real prescription crops (Dataset 3 text labels)}
\label{sec:res-rx3}
\begin{figure}[t]
\centering
\includegraphics[width=\columnwidth]{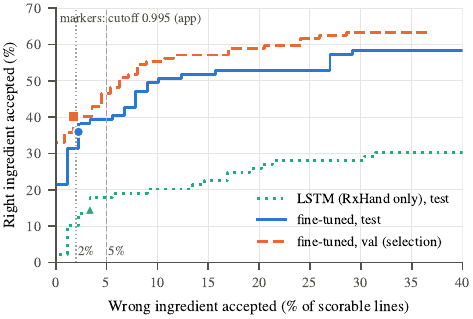}
\caption{Risk--coverage on real prescription crops. The val curve (orange) is where the cutoff was selected; markers: $\tau_a=0.995$. With only 89 scorable test lines, one wrong acceptance moves WA by 1.1 points.}
\label{fig:rx3}
\end{figure}

On the 165 test crops, 21 are unreadable and 23 are not medicines; of the 121 medicine lines, 32 (26\%) name a brand that is not in the dictionary (abbreviations, misspellings by the prescriber, or brands absent from both registries) and can never be matched. Table~\ref{tab:rx3} and Fig.~\ref{fig:rx3} report the remaining 89 scorable lines. Before fine-tuning, the reader's CER on medicine lines was 0.63 (versus 0.26 on RxHandBD), a large domain gap: real crops contain strengths and brackets, capitals, ruled lines and neighbouring strokes. Fine-tuning on the train crops reduced CER to 0.40 and roughly doubled the correct-acceptance rate at every cutoff (paired McNemar at 0.995: 23 lines correct only after fine-tuning, 4 only before, $p=3.1\times10^{-4}$), while performance on RxHandBD was unchanged (CER 0.257 vs 0.256).

\begin{table}[t]
\caption{Real prescription crops, test split: 89 scorable medicine lines (IC/WA, \% with 95\% CI). Last rows: app behaviour over all 165 boxes at $\tau_a{=}0.995$, $\tau_s{=}0.9$.}
\label{tab:rx3}
\centering
\renewcommand{\arraystretch}{1.1}
\begin{tabular}{@{}lcc@{}}
\toprule
 & \textbf{RxHand-only} & \textbf{Fine-tuned} \\
\midrule
CER (121 medicine lines) & 0.63 & 0.40 \\
Brand part read exactly & 4.1\ci{1.8}{9.3} & 16.5\ci{11.0}{24.2} \\
Top-1 IC (no cutoff) & 33.7\ci{24.7}{44.0} & 58.4\ci{48.0}{68.1} \\
IC @0.98 (5\% cap on val) & 18.0\ci{11.4}{27.2} & 44.9\ci{35.0}{55.3} \\
WA @0.98 & 5.6\ci{2.4}{12.5} & 7.9\ci{3.9}{15.4} \\
\textbf{IC @0.995 (2\% cap, deployed)} & 14.6\ci{8.7}{23.4} & \textbf{36.0}\ci{26.8}{46.3} \\
\textbf{WA @0.995} & 3.4\ci{1.2}{9.4} & \textbf{2.2}\ci{0.6}{7.8} \\
IC / WA @0.9999 & 9.0 / 1.1 & 23.6 / 1.1 \\
\midrule
Auto-accepted (of 165 boxes) & 13.3\% & 27.9\% \\
``Is this X?'' (of 165 boxes) & 13.3\% & 24.8\% \\
\bottomrule
\end{tabular}
\end{table}

\subsection{Simulated use: time per patient}
\label{sec:res-time}
\begin{figure}[t]
\centering
\includegraphics[width=\columnwidth]{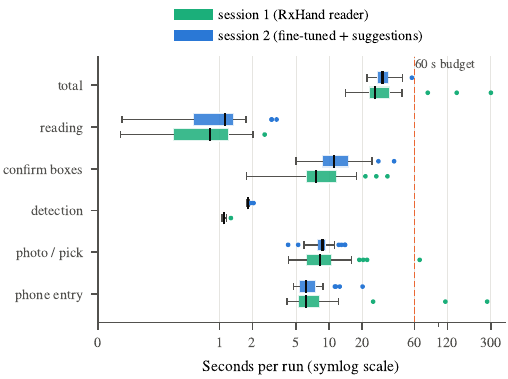}
\caption{Per-step time per run in both sessions (symlog axis). Machine steps (detection, reading) take about a second; human steps dominate.}
\label{fig:time}
\end{figure}

\begin{figure}[t]
\centering
\includegraphics[width=\columnwidth]{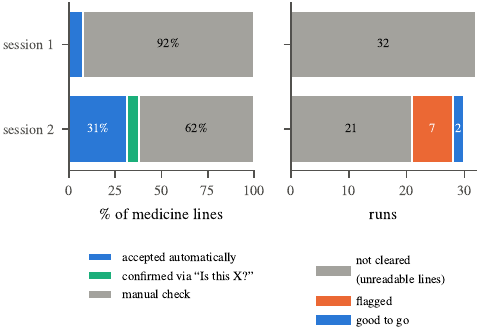}
\caption{Left: disposition of medicine lines per session. Right: run outcomes. ``Not cleared'' means at least one line was unreadable, so the app refused to say ``Good to go''.}
\label{fig:sessions}
\end{figure}

The author processed the 30 Dataset~3 test prescriptions twice, with distinct dummy phone numbers, acting as the physician: session~1 (32 runs: 25 gallery, 7 camera) with the RxHand-only reader at 0.9999 and no confirmation tier; session~2 (30 runs, all gallery) with the fine-tuned reader at 0.995 and ``Is this X?''. Table~\ref{tab:time} and Figs.~\ref{fig:time}--\ref{fig:sessions} summarise the logs.

\begin{table}[t]
\caption{Per-run timing from the application logs (median with 95\% bootstrap CI, seconds) and line dispositions. S1/S2 = session 1/2.}
\label{tab:time}
\centering
\renewcommand{\arraystretch}{1.1}
\scriptsize
\setlength{\tabcolsep}{2.5pt}
\begin{tabular}{@{}lcc@{}}
\toprule
 & \textbf{S1} ($n{=}32$) & \textbf{S2} ($n{=}30$) \\
\midrule
Phone entry & 6.2\ci{5.6}{7.2} & 6.2\ci{5.6}{7.4} \\
Photo / pick & 8.2\ci{6.7}{8.7} & 8.7\ci{8.1}{9.4} \\
Detection & 1.1\ci{1.1}{1.1} & 1.8\ci{1.8}{1.8} \\
Confirm boxes & 7.6\ci{6.5}{8.3} & 11.1\ci{9.6}{13.6} \\
Reading (all lines) & 0.9\ci{0.8}{1.1} & 1.1\ci{0.9}{1.3} \\
\textbf{Total per patient} & \textbf{26.4}\ci{24.0}{33.3} & \textbf{30.7}\ci{28.3}{33.7} \\
90th percentile / max & 45.9 / 299.6 & 44.1 / 56.9 \\
Machine / human (s) & 2.2 / 24.3 & 3.1 / 27.9 \\
Runs $\leq$ 60\,s & 29/32 (90.6\%) & 30/30 (100\%) \\
\midrule
Medicine lines & 146 & 153 \\
Manual check & 92.5\%\ci{87.0}{95.7} & 62.1\%\ci{54.2}{69.4} \\
Boxes det./removed/drawn & 115 / 5 / 49 & 134 / 6 / 33 \\
Suggestions shown/yes/no & --- & 28 / 10 / 18 \\
Unreadable/flagged/good & 32 / 0 / 0 & 21 / 7 / 2 \\
\bottomrule
\end{tabular}
\end{table}

All 62 runs were offline; the application computed for a median of 2--3\,s per patient, the rest being human time. Total time did not differ significantly between sessions (Mann Whitney $p=0.16$), but confirming boxes took longer in session~2 ($p=0.005$), consistent with more lines to inspect and answering suggestions (median 9.2\,s to answer all suggestions in a run). The share of lines left for manual check fell from 92.5\% to 62.1\% (Fisher $p=1.9\times10^{-10}$), and only session~2 produced any actionable result: 7 flagged runs (14 interaction warnings) and 2 ``Good to go''. Of the three session-1 runs over 60\,s, two were idle time before typing the phone number (277\,s and 115\,s) and one a 67\,s gallery pick; none was computation. Detection took 1.8\,s in session~2 versus 1.1\,s in session~1 with the same model; we did not identify the cause (thermal state is one candidate).

\subsection{Qualitative error analysis}
\label{sec:qual}
\begin{figure*}[t]
\centering
\includegraphics[width=\textwidth]{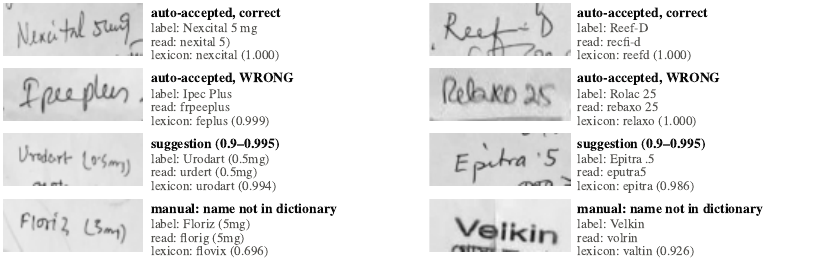}
\caption{Real test-split crops with the author's label, the reader's greedy text, and the lexicon decision with its ingredient-set posterior. ``Auto-accepted, WRONG'' rows are the safety-relevant failures; ``Rolac 25'' $\rightarrow$ ``Relaxo'' is the same line the author wrongly confirmed in Fig.~\ref{fig:screens}c.}
\label{fig:crops}
\end{figure*}

We reviewed all 60 screenshots of session~2 and the reading failures on the test crops (examples in Fig.~\ref{fig:crops}). Recurring failure modes:
\begin{enumerate}
\item \textbf{Confident wrong acceptance.} ``Napa'' syrup was accepted as ``MSL 10mg/5ml'' (morphine), generating three spurious morphine interaction warnings; ``Ipec Plus'' was accepted as ``Feplus''. Two of 89 scorable test lines were accepted wrongly at the deployed cutoff.
\item \textbf{Wrong human confirmation.} The author answered \emph{Yes} to ``Relaxo (dantrolene)'' for a line labelled ``Rolac 25'' (ketorolac), producing a spurious dantrolene--pregabalin warning (Fig.~\ref{fig:screens}c). Of 28 suggestions, 18 were rejected; the confirmation step catches most but not all wrong guesses and may anchor the user.
\item \textbf{Unknown names.} 26\% of written medicine names are not dictionary brands; these always fall to manual check.
\item \textbf{Alert fatigue.} Many warnings involved low-importance pairs (e.g.\ pantoprazole + vitamin D3, rosuvastatin + calcium carbonate). Without severity data the system cannot suppress them, contrary to the severity-gated design of our earlier version~\cite{poly2018}.
\item \textbf{Ingredient unions.} Brands registered with several formulations resolve to the union of ingredients (e.g.\ ``Xinc'' $\rightarrow$ five ingredients; ``Napa'' includes caffeine because one registered variant does), which can over-flag.
\item \textbf{Detector misses} required the user to draw 49 and 33 boxes in the two sessions.
\item \textbf{Long confirmation prompts} for multivitamin brands with 30+ ingredients are hard to read.
\end{enumerate}

\subsection{Functional test: cross-visit medication history}
\label{sec:crossvisit}
\begin{figure}[t]
\centering
\includegraphics[width=0.32\columnwidth]{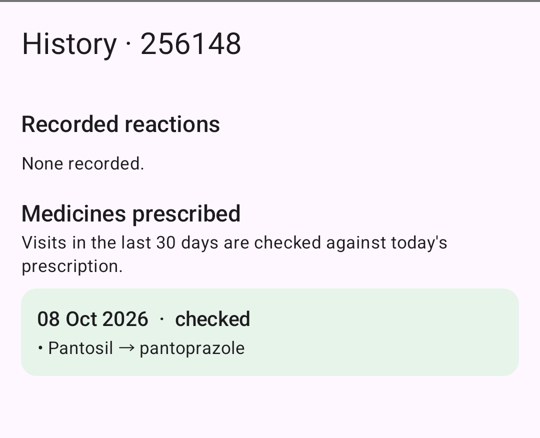}\hfill
\includegraphics[width=0.32\columnwidth]{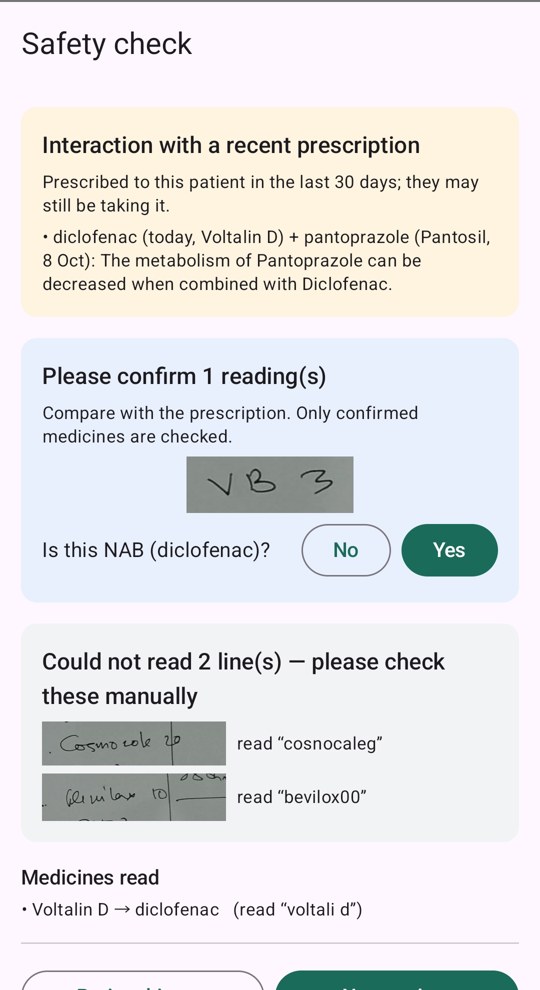}\hfill
\includegraphics[width=0.32\columnwidth]{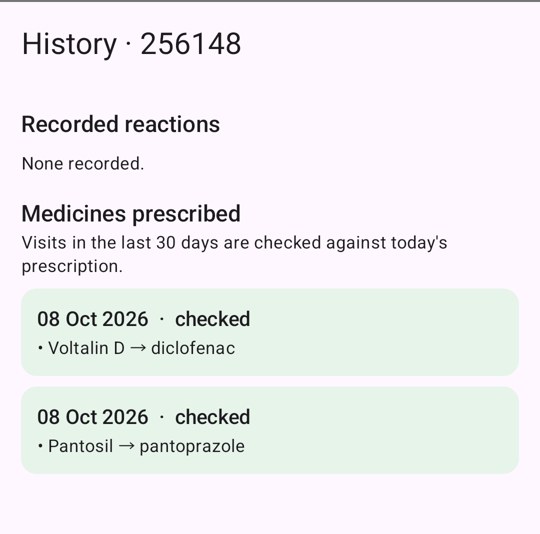}
\caption{Cross-visit functional test. Left: history after visit~1 (Pantosil stored). Middle: visit~2 flags diclofenac (today) + pantoprazole (visit~1, ``8 Oct''); the rejected ``NAB'' suggestion is not stored. Right: both visits listed as inside the 30-day window.}
\label{fig:crossvisit}
\end{figure}

The cross-visit check was implemented after the two evaluated sessions, so it is \emph{not} reflected in any number above. It was verified in two ways. First, seven instrumented tests on the phone, using an in-memory database: a clopidogrel prescription 3 or 30 days earlier is flagged against today's omeprazole, one 40 days earlier is not; another patient's history, the current visit itself and a non-interacting earlier medicine raise nothing; and a pair that is entirely within today's prescription is reported once, by check B. All seven passed. Second, one end-to-end functional run by the author with a dummy patient and two Dataset~3 test prescriptions (Table~\ref{tab:crossvisit}, Fig.~\ref{fig:crossvisit}).

\begin{table}[t]
\caption{Functional test of the cross-visit check: one dummy patient, two visits on the same day (application log and stored history).}
\label{tab:crossvisit}
\centering
\scriptsize
\setlength{\tabcolsep}{3pt}
\begin{tabular}{@{}>{\raggedright\arraybackslash}p{2.0cm}>{\raggedright\arraybackslash}p{2.8cm}>{\raggedright\arraybackslash}p{3.1cm}@{}}
\toprule
 & \textbf{Visit 1} & \textbf{Visit 2} \\
\midrule
History before check & none stored & visit 1 (in window) \\
Read automatically & Pantosil $\rightarrow$ pantoprazole & Voltalin~D $\rightarrow$ diclofenac \\
Manual / ``Is this X?'' & 2 manual & 2 manual; ``NAB (diclofenac)?'' $\rightarrow$ No \\
Result & not cleared (unreadable) & \textbf{flagged}: diclofenac (today) + pantoprazole (visit 1) \\
Stored for later visits & Pantosil & Voltalin~D (NAB not stored) \\
Total time & 200.6\,s & 52.9\,s \\
\bottomrule
\end{tabular}
\end{table}

The run shows the mechanism working as designed: the application stored the resolved medicines but not the rejected suggestion, and it checked the second visit against the first. The run also shows the same alert-fatigue limitation as check B: the flagged pair is a documented metabolic interaction of low clinical importance, and without severity data the application shows it as prominently as a serious one. A single dummy patient demonstrates function, not usefulness. We do not know how often the check catches a clinically relevant pair among real returning patients.

\subsection{What was not shown}
We implemented the reaction-history path, which reads a written note such as ``Allergic: Ceftriaxone'' with ML Kit, stores the ingredient against the phone number, and flags it at a later visit, but no logged run exercised it. We can therefore answer RQ3 only for interactions within a prescription and, in one functional run, across visits. No run involved a real site prescription, a clinician, or a real returning patient.

\section{Discussion}
\subsection{Is the system feasible?}
\textbf{Workflow: yes, under simulated use.} All 30 runs in the final configuration finished within 60\,s (95\% CI 88.6--100\%), and on-device computation took about 3\,s. Human handling, not the phone, sets the pace. \textbf{Retrieval: yes.} Once the application has the ingredients, the interaction lookup is instantaneous. The lookup flagged 7 of 30 session-2 prescriptions, but a wrong reading or a wrong confirmation caused at least two of these flags (Sec.~\ref{sec:qual}). \textbf{Reading: only partially.} Even after fine-tuning, the physician must still read or confirm about two-thirds of real medicine lines, and the reader still accepts 2\% of scorable lines wrongly. The tool in this form resolves a minority of lines automatically and makes the physician look at the rest, so it works as an assistant, not as an autonomous checker.

\subsection{What made the difference}
Three choices, not model architecture, produced the largest gains. First, we constrained output to registered brands: lexicon decoding raised top-1 ingredient accuracy on Dataset~4 from 30.0\% with ML Kit to 54--64\%. Second, we measured confidence at the level the safety check uses, the ingredient set. Third, we added a small amount of in-domain labelled data (660 crops), which halved the real-crop error rate where public word datasets did not transfer. Protocol discipline mattered as much: cutoffs chosen on the benchmark overstated acceptable accuracy by roughly 15 percentage points (38.0\% vs 22.9\% for the same model).

\subsection{Threats to validity and limitations}
\label{sec:threats}
\begin{itemize}
\item \textbf{Author as only user and only annotator.} Timing reflects a user who built the tool and knew the test images; real physicians may be slower or faster. Labels were single-coded.
\item \textbf{Gallery images, not live capture.} Session~2 used only stored photographs; camera conditions (glare, blur, angle) were tested in 7 runs only.
\item \textbf{Small real-crop test set.} 89 scorable lines give wide intervals (WA 2.2\% [0.6, 7.8]).
\item \textbf{Post-hoc cap change.} The 2\% cap was adopted after the 5\% cap failed on test; both operating points are reported, but the deployed cutoff is not a clean pre-registered choice.
\item \textbf{Single dataset family.} Datasets~3--5 come from Bangladeshi prescriptions of unknown provenance and licence~\cite{ds_rxhandbd,ds_d4,ds_rxboxes}; generalisation to the study hospital is untested. Dataset~3 also contains boxes on non-medicines.
\item \textbf{Knowledge-base limits.} No interaction severity; 48\% of dictionary ingredients absent from the DDI table; synonym coverage is hand-curated; ingredient unions over-flag.
\item \textbf{No outcome, usability or identity evaluation} (Table~\ref{tab:plan}); no ethics review was required because no human data were collected, but any field study will need one.
\item \textbf{Cross-visit history is only as complete as the reading.} Lines that were unreadable or rejected are never stored, so a returning patient's history omits exactly the medicines the reader failed on; the 30-day window is a fixed assumption about how long a medicine may still be taken.
\item \textbf{Licensing.} YOLOv8 weights are AGPL-3.0~\cite{jocher2023yolov8}; the DDI table is DrugBank-derived~\cite{wishart2018drugbank} and its licence must be checked before redistribution.
\end{itemize}

\section{Conclusion and Future Work}
We built an offline smartphone pipeline that photographs a handwritten Bangladeshi prescription, resolves each medicine to its active ingredients, and checks those ingredients against documented drug--drug interactions and the patient's recorded reactions. We measured each component. The workflow fits within a one-minute consultation in simulated use, and retrieval works once the application has the ingredients. Reading the handwriting is the bottleneck. With lexicon-constrained decoding and modest in-domain fine-tuning, the reader resolves about a third of real medicine lines automatically, wrongly accepting roughly 2\%; it turns a quarter into one-tap confirmations and leaves the rest to the physician. After the evaluation, we added a 30-day cross-visit medication history, which extends the interaction check to a returning patient's recent prescriptions. It passed seven automated tests and one two-visit functional test, but we have not evaluated it.

In order of expected value, we plan to: (1) run a field study with physicians at the target upazila health complex, with ethics approval, to measure time, usability, and acceptance on live capture; (2) label more in-domain prescription crops, with two annotators per crop; (3) add interaction severity data to suppress low-value alerts, which now matters for both within-prescription and cross-visit checks; (4) evaluate the cross-visit history with real returning patients and replace the fixed 30-day window with a per-drug duration; (5) evaluate the reaction-note path and the re-linking of returning patients by phone number; and (6) test retrospectively on de-identified site cases with documented reactions.

\bibliographystyle{IEEEtran}
\bibliography{refs}

\end{document}